\documentclass [aip,apl,a4paper,reprint,superscriptaddress] {revtex4-1}
\usepackage{amsmath}
\usepackage{graphicx}
\usepackage{subfigure}
\usepackage{xcolor}

\begin{document}
\title{Controlling shedding characteristics of condensate drops using electrowetting}

\author{Ranabir Dey} 
\affiliation{Physics of Complex Fluids, MESA+ Institute for Nanotechnology, University of Twente, PO Box 217, 7500 AE Enschede, The Netherlands}

\author{Jander Gilbers} 
\affiliation{Physics of Complex Fluids, MESA+ Institute for Nanotechnology, University of Twente, PO Box 217, 7500 AE Enschede, The Netherlands}

\author{Davood Baratian} 
\affiliation{Physics of Complex Fluids, MESA+ Institute for Nanotechnology, University of Twente, PO Box 217, 7500 AE Enschede, The Netherlands}

\author{Harmen Hoek} 
\affiliation{Physics of Complex Fluids, MESA+ Institute for Nanotechnology, University of Twente, PO Box 217, 7500 AE Enschede, The Netherlands}

\author{Dirk van den Ende}
\affiliation{Physics of Complex Fluids, MESA+ Institute for Nanotechnology, University of Twente, PO Box 217, 7500 AE Enschede, The Netherlands}

\author{Frieder Mugele}
\email[e-mail:]{f.mugele@utwente.nl}
\affiliation{Physics of Complex Fluids, MESA+ Institute for Nanotechnology, University of Twente, PO Box 217, 7500 AE Enschede, The Netherlands}

\begin{abstract}
We show here that ac electrowetting (ac-EW) with structured electrodes can be used to control the gravity-driven shedding of drops condensing onto flat hydrophobic surfaces. Under ac-EW with straight interdigitated electrodes, the condensate drops shed with relatively smaller radii due to the ac-EW-induced reduction of contact angle hysteresis. The smaller shedding radius, coupled with the enhanced growth due to coalescence under EW, results in increased shedding rate. We also show that the condensate droplet pattern under EW can be controlled, and the coalescence can be further enhanced, using interdigitated electrodes with zigzag edges. Such enhanced coalescence in conjunction with the electrically-induced trapping effect due to the electrode geometry results in larger shedding radius, but lower shedding rate. However, the shedding characteristics can be further optimized by applying the electrical voltage intermittently. We finally provide an estimate of the condensate volume removed per unit time in order to highlight how it is enhanced using ac-EW-controlled dropwise condensation.    
\end{abstract}
\maketitle

Dropwise condensation is important in a wide range of technologies like water-harvesting systems \cite{milani2011evaluation}, desalination systems \cite{khawaji2008advances}, and heat exchangers \cite{kim2002air}. The effectiveness of all these technologies depends on the efficient volumetric collection rate of the condensate, and hence, depends on the shedding of the condensate drops from the condensing surface. The continuous drop shedding exposes bare surface for renewed nucleation and growth of the condensate drops culminating in efficient vapour-to-liquid phase change and enhanced condensate collection \cite{rose2002dropwise}. To this end, the enhanced mobility and shedding of condensate drops have been studied on superhydrophobic nanostructured surfaces \cite{boreyko2009self, miljkovic2012effect, miljkovic2012jumping, miljkovic2013electric}, wettability-patterned surfaces \cite{ghosh2014enhancing}, liquid impregnated textured surfaces \cite{anand2012enhanced, tsuchiya2017liquid}, and biomimetic surfaces \cite{park2016condensation}. All these approaches towards enhancing droplet mobility are passive in nature, relying solely on the topographical and/or chemical patterning of the condensing surface. As an alternative, recently we have demonstrated that an alternating (ac) electric field in an electrowetting (EW) configuration with structured electrodes can be used to actively control the mobility of condensate drops on homogeneous hydrophobic surfaces \cite{baratian2018breath}. The growth of the condensate drops under EW is characterized by their migration to the size-dependent locations of the minima in the corresponding electrostatic energy landscapes and by enhanced coalescence \cite{baratian2018breath}. The use of electrical forces to control condensate droplet pattern (breath figures) evolution is-- to our knowledge-- a completely new approach. While in our previous study we focused on the evolution and statistics of the condensate droplet pattern, the present work is devoted to the analysis of subsequent gravity-driven shedding of condensate drops under the influence of ac-EW. Such a study is essential for the effective implementation of EW for technological applications involving dropwise condensation.

In this letter, we demonstrate that the gravity-driven shedding characteristics of condensate drops can be indeed controlled using ac-EW with structured electrodes. In general, a condensate drop on a vertical substrate begins to shed under gravity only when the drop reaches a certain critical `shedding' radius $R_{sh}$ at which its weight overcomes the inherent contact angle hysteresis force \cite{de2004capillarity}. It has been also demonstrated that ac-EW in air results in the reduction of the effective contact angle hysteresis culminating in enhanced mobilization of sessile drops \cite{li2008make, t2011electrically}. We show here that under ac-EW with straight interdigitated electrodes, the underlying reduction in effective contact angle hysteresis and the enhanced coalescence-dominated growth result in smaller $R_{sh}$ and increased shedding rate $(f_{sh})$ of the condensate drops, as compared to the classical no EW case. Interestingly, the shedding characteristics under ac-EW can be further altered using interdigitated electrodes with zigzag edges. In this case, the enhanced mobility of condensate drops due to the non-uniform electrical force distribution and the eventual electrical trapping effect result in larger $R_{sh}$ and lower $f_{sh}$; however, the overall condensate removal rate increases. Finally, we demonstrate that the shedding of condensate is further enhanced by applying the electrical voltage intermittently instead of continuously. 

\begin{figure} [ht]
	\includegraphics*[width=\columnwidth]{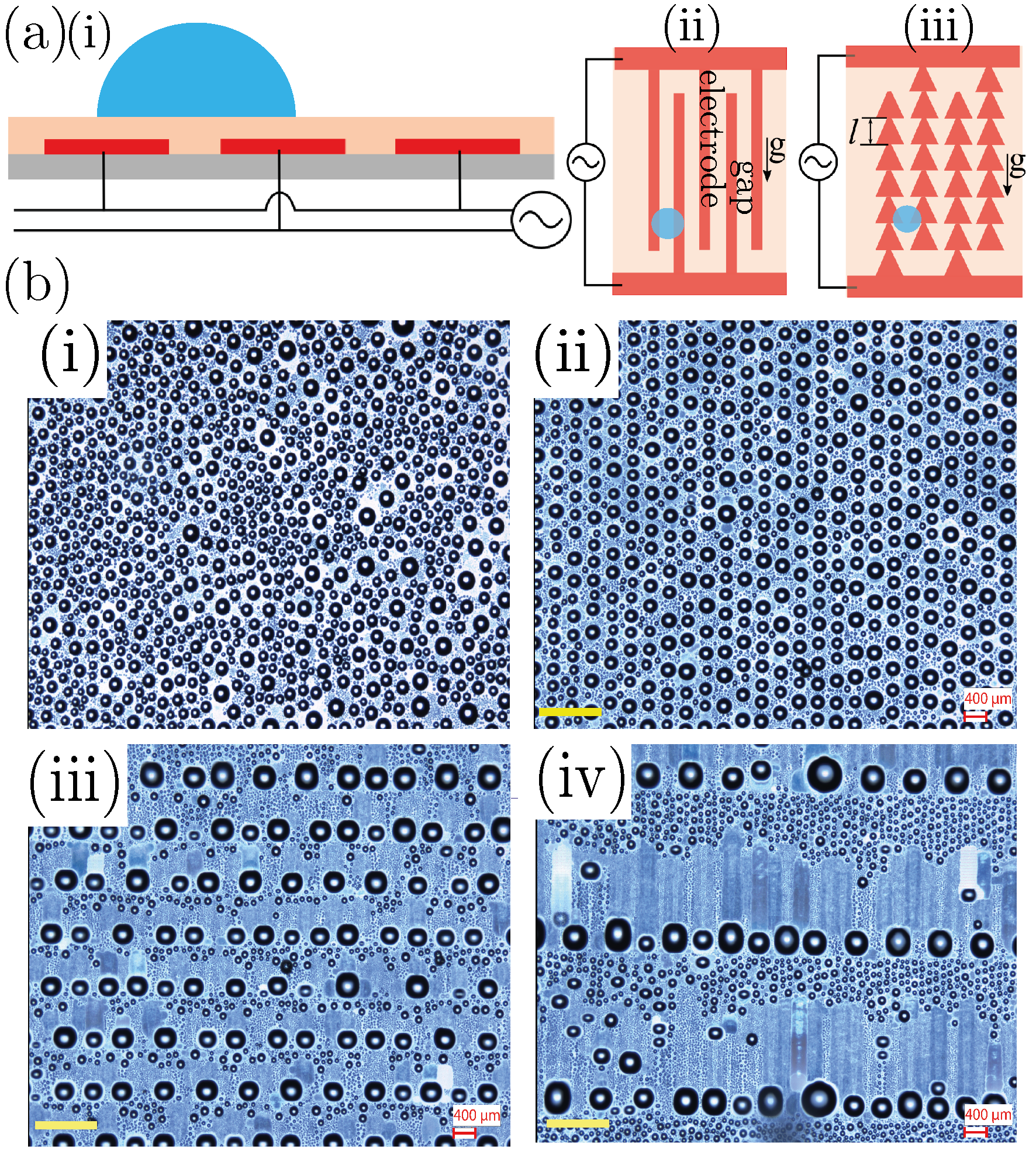}
	\caption{\label{Fig. 1} (a)(i) Schematic of the substrate used for the condensation experiments. The substrate consists of transparent interdigitated ITO electrodes (red) patterned on a glass substrate (gray), which is coated with a hydrophobic dielectric film (orange). Schematics of the interdigitated electrode (electrode-gap) designs are also shown here-- (a)(ii) straight interdigitated electrodes, (a)(iii) zigzag interdigitated electrodes; the distance $l$ between the consecutive triangular elements is varied to create different electrode designs. (b) Comparison between condensate droplet patterns (at approximately the same time instant) (i) without EW and under EW ($U_{rms}=150$ V; $f=1$ kHz) with different electrode designs-- (ii) straight interdigitated electrodes, (iii) zigzag interdigitated electrodes with $l=1000$ $\mu$m and (iv) $l=3000$ $\mu$m. Gravity points from top to bottom along the electrodes. The yellow bars in (b) represent 1 mm.}
\end{figure}

The experimental setup is identical to that used in our previous study \cite{baratian2018breath} (see Sec. S1 in the supplemental material for a schematic). The condensing substrate consists of interdigitated ITO electrodes on a glass substrate, which is coated with a hydrophobic dielectric film-- 2 $\mu$m thick Parylene C layer topped with an ultrathin layer of Cytop (Fig. \ref{Fig. 1}(a)(i)). For the straight interdigitated electrodes (Fig. \ref{Fig. 1}(a)(ii)) the width of both the electrodes and the gaps is 200 $\mu$m; for the zigzag interdigitated electrodes (Fig. \ref{Fig. 1}(a)(iii)) the base and the apex of each triangular element for both electrodes and gaps are 250 $\mu$m and 50 $\mu$m wide respectively, while the distance $l$ between the consecutive triangular elements is varied from 500 $\mu$m to 3000 $\mu$m to create different electrode designs. For EW, an ac voltage with a frequency of $f=1$ kHz and a maximum magnitude of 150 V $U_{rms}$ is applied across the electrodes (Fig. \ref{Fig. 1}(a)). Thereafter, a stream of vapour-air mixture at a flow rate of 3.6 l/min and a temperature of 41.8 $^\circ$C is passed through a condensation chamber, in which the substrate is maintained at a temperature of 11.5 $^\circ$C. Condensation experiments are performed without EW, and under ac-EW by applying different magnitudes of $U_{rms}$ using straight and zigzag interdigitated electrodes with $l=$500 $\mu$m, 1000 $\mu$m, and 3000 $\mu$m. The condensation process, including the shedding events, is monitored for $\sim5$ minutes using a high resolution camera.

 In absence of EW ($U_{rms}=0$ V), the condensate drops are apparently randomly distributed with smaller average sizes (Fig. \ref{Fig. 1}(b)(i); Movie S1 in the supplementary material). In contrast, under ac-EW with straight interdigitated electrodes, the condensate drops with diameters comparable to the gap width are aligned along the corresponding electrostatic energy minima locations at the gap centres (Fig. \ref{Fig. 1}(b)(ii); Movie S2). As discussed in our earlier study \cite{baratian2018breath}, this alignment process is accompanied by a sharp increase of the average drop size. The latter is caused by the cascades of coalescence events triggered by the EW-induced migration of the drops. In this study, the underlying fact that the coalescence-induced growth of the condensate drops under EW can be further enhanced by moving the drops in a particular direction using non-uniform electrical forces motivated the use of the zigzag interdigitated electrodes towards altering the final shedding characteristics. The converging gap geometry results in a net downward electrical force on a condensate drop which moves it towards the gap apex. Such sweeping of condensate drops results in enhanced coalescence culminating in increased growth of average drop size. However, the droplets thus mobilized eventually accumulate at the apices of the triangular gap elements due to the electrical trapping effect at these locations (schematic in Fig. \ref{Fig. 1}(a)(iii)). Hence, the condensate droplet pattern under EW with zigzag interdigitated electrodes also has a periodicity along the electrodes which is given by $l$ (compare Fig. \ref{Fig. 1}(b)(ii) and Fig. \ref{Fig. 1}(b)(iii) or (iv)). For longer $l$, the condensate drops sweep a longer distance on the condensing surface resulting in larger sizes of the trapped condensate droplets and also longer periodicity along the electrodes (compare Fig. \ref{Fig. 1}(b)(iii) and Fig. \ref{Fig. 1}(b)(iv); Movie S3, Movie S4 and Movie S5 show the condensate droplet pattern evolution corresponding to $l=$1000 $\mu$m, 3000 $\mu$m, and 500 $\mu$m respectively).

\begin{figure} [ht]
	\includegraphics*[width=\columnwidth]{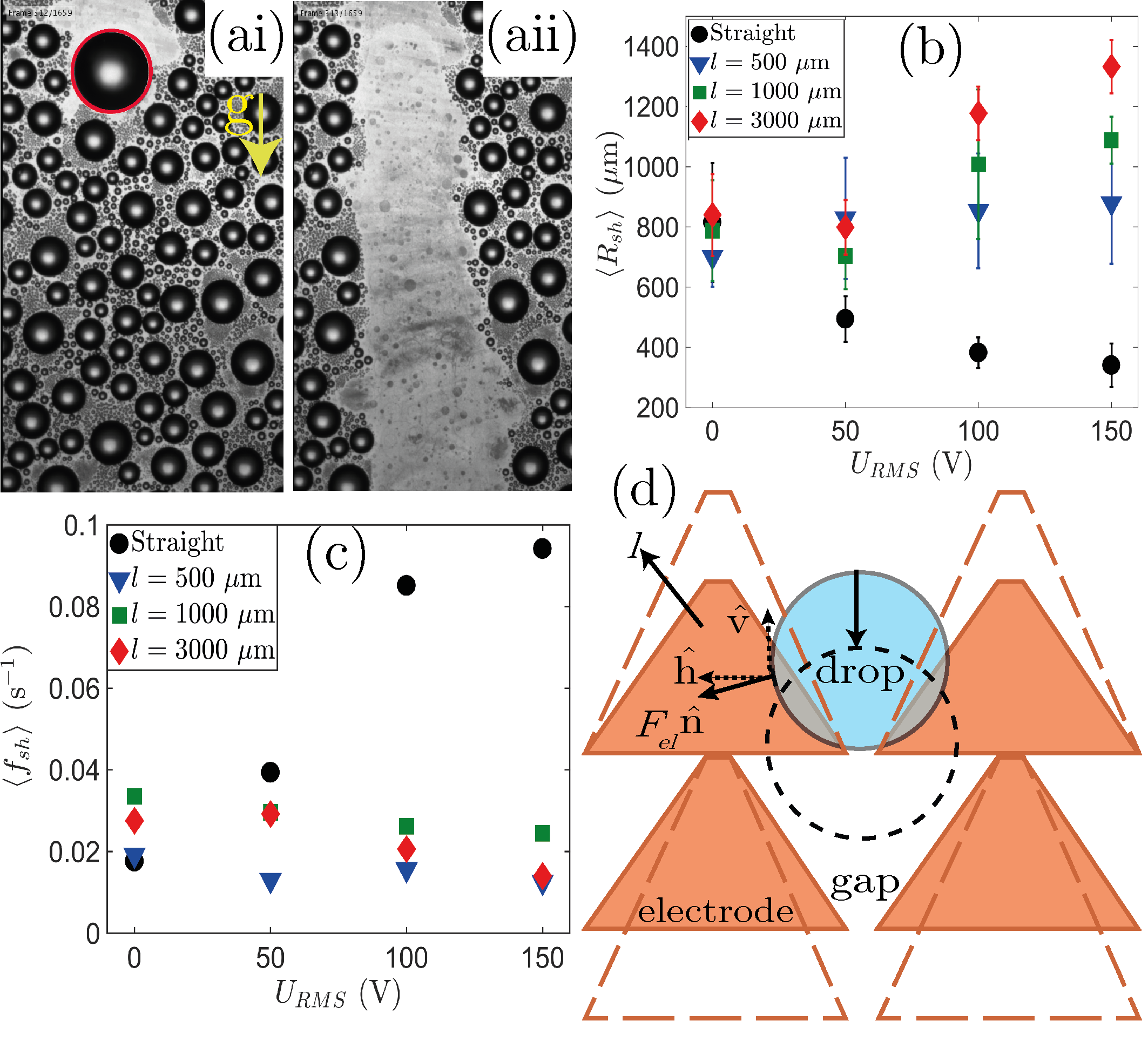}
	\caption{\label{Fig. 2} (ai) Representative image of a condensate drop (outlined in red) about to shed under gravity. The radius of the drop at this instant is defined as the shedding radius $R_{sh}$. (aii) The consequent frame showing the clearing of the surface due to the shedding of the same droplet. (b) Variations of the average droplet shedding radius $\langle R_{sh} \rangle$ with the applied voltage $U_{rms}$ for different electrode designs, i.e. straight (Fig. \ref{Fig. 1}(a)(ii)) and zigzag interdigitated electrodes with different values of $l$ (Fig. \ref{Fig. 1}(a)(iii)). (c) Variations of the average shedding rate $\langle f_{sh} \rangle$ with $U_{rms}$ for the different electrode designs. (d) Schematic of a condensate droplet during dropwise condensation under ac-EW with zigzag interdigitated electrodes of varying $l$. The schematic is not to scale.}
\end{figure}

 The final gravity-driven shedding characteristics are quantified here by the average shedding radius $\langle R_{sh} \rangle$ of the condensate drops, and the average shedding rate $\langle f_{sh} \rangle$. $\langle R_{sh} \rangle$ represents the average value of $R_{sh}$ evaluated using on an average more than 10 shedding events corresponding to a particular EW condition, where $R_{sh}$ is evaluated using an image analysis procedure (Fig. \ref{Fig. 2}(a); see Sec. S1 in the supplementary material). $\langle f_{sh} \rangle$ is evaluated by dividing the total number of recorded shedding events with the total time required for these starting from the opening of the vapour valve for the initiation of the condensation process. Fig. \ref{Fig. 2}(b) clearly shows that $\langle R_{sh} \rangle$ progressively reduces under ac-EW, with straight interdigitated electrodes, with increasing $U_{rms}$; $\langle R_{sh} \rangle$ for $U_{rms}=150$ V is approximately 50\% smaller than that observed without EW. Note that the values of $R_{sh}$ are typically larger than the electrode pitch and hence the shedding drops cover a few electrodes. In absence of EW, considering the balance of the droplet weight and the inherent contact angle hysteresis (CAH) force \cite{de2004capillarity}, $\langle R_{sh} \rangle$ can be estimated as $\langle R_{sh} \rangle \approx \sqrt{\frac{3}{\pi}}\lambda_c (\Delta \cos \theta)^{1/2}$, where $\lambda_c=\sqrt{\frac{\gamma}{\rho g}}$ is the capillary length, $\gamma$ and $\rho$ are the surface tension and the density of water respectively, and $\Delta \cos \theta$ is the difference between the cosines of the receding and advancing contact angles of water drops on the condensing surface which gives a measure of the involved CAH (see Sec. S2 in the supplementary material). In regard to CAH, it is now established that under ac-EW in air the effective CAH gradually decreases with increasing ac voltage \cite{li2008make, t2011electrically, gao2018contact}. This reduction in CAH is due to the oscillatory electrical force induced depinning of the three-phase contact line from the random surface heterogeneities, and can be expressed as $\Delta \cos \theta(U_{rms}) \approx \Delta \cos \theta_0- \alpha \beta U_{rms}^2$, where $\Delta \cos \theta_0$ is the value of $\Delta \cos \theta$ for $U_{rms}=0$ V, $\beta$ is the ratio of the effective dielectric capacitance per unit area and $\gamma$, and $\alpha$ is a coefficient characterizing the efficiency of the ac-EW induced CAH reduction mechanism (generally $\alpha \sim 1$) \cite{li2008make, t2011electrically}. In this way, ac-EW reduces $\langle R_{sh} \rangle$ with $U_{rms}$ for the case of straight interdigitated electrodes (Fig. \ref{Fig. 2}(b)); hence, the corresponding voltage dependent shedding radius can be estimated as $\langle R_{sh} \rangle \approx \sqrt{\frac{3}{\pi}}\lambda_c [\Delta \cos \theta_0 - \alpha \beta U_{rms}^2]^{1/2}$ (also see Sec. S2 in the supplementary material). It must be noted here that CAH does not go on decreasing with increasing $U_{rms}$, but stabilizes at a finite, albeit smaller, value at moderate values of $U_{rms}$ \cite{li2008make,t2011electrically}. Accordingly, the reduction in $\langle R_{sh} \rangle$ is insignificant for higher values of $U_{rms}$ (Fig. \ref{Fig. 2}(b); Sec. S2 in the supplementary material). The gradually decreasing value of $\langle R_{sh} \rangle$, coupled with the enhanced coalescence induced droplet growth under EW, results in increasing value of $\langle f_{sh} \rangle$ (Fig. \ref{Fig. 2}(c)). In case of zigzag interdigitated electrodes, the increasing non-uniform overlap area between a condensate droplet footprint and the active electrode elements results in a net electrical force on the droplet in the direction of the converging gap (Fig. \ref{Fig. 2}(d)). This force is obtained by integrating the vertical $(\hat{v})$ component of the normal electrical force per unit length $(\vec{F_{el}} \approx \beta \gamma U_{rms}^2 \hat{n})$ along the droplet contact line length on top of the active electrode elements (Fig. \ref{Fig. 2}(d)); also see Sec. S3 in the supplementary material). This electrical force sweeps the condensate drops towards the gap apices thereby enhancing coalescence and droplet growth. However, at the gap apex the orientation of the net vertical electrical force reverses direction (from downward to upward) due to the electrode geometry, which consequently traps the droplet at that location (dotted circle in Fig. \ref{Fig. 2}(d)), in a manner similar to droplet trapping by electrically tunable defects \cite{ghosh2014trapping}. The condensate drops thus accumulate at the gap apices till the droplet weight overcomes the contact angle hysteresis force and the additional electrical trapping force. So, for this case $\langle R_{sh} \rangle$ can be estimated from the relation 
 \begin{equation}
   \langle R_{sh} \rangle^3 \approx \frac{3}{\pi} \lambda_c^2 (\Delta \cos \theta_0 - \alpha \beta U_{rms}^2)\langle R_{sh} \rangle+ \frac{1.5}{\pi} \lambda_c^2\beta U_{rms}^2 \Delta l_c^h \label{eq. 1}
 \end{equation}
 where the last term in Eq. \ref{eq. 1} is due to the additional electrical trapping force, $\Delta l_c^h$ represents the difference between the horizontal projections of the total droplet contact line length on the zigzag electrode elements above the horizontal droplet footprint diameter and of the same below it, and $\alpha$ here takes care of the possible non-uniformity in the ac-EW-induced CAH reduction mechanism due to the zigzag electrode geometry (also see Sec. S3 in the supplementary material). Note that $\Delta l_c^h<2 \langle R_{sh} \rangle$. Considering $2\langle R_{sh} \rangle$ as a scale for $\Delta l_c^h$, it can be inferred from Eq. \ref{eq. 1} that for small values of $U_{rms}$ the reduction in CAH force and the additional electrical trapping force (last two terms on RHS in Eq. \ref{eq. 1}) almost balance each other. Consequently, $\langle R_{sh} \rangle$ for zigzag interdigitated electrodes remains relatively unchanged for small values of $U_{rms}$ (Fig. \ref{Fig. 2}(b)). However,  for large values of $U_{rms}$, the CAH force remains constant at a small finite value, while the magnitude of the electrical trapping force progressively increases $(\propto U_{rms}^2)$. Hence, $\langle R_{sh} \rangle$ increases with higher values of $U_{rms}$ for zigzag interdigitated electrodes (Fig. \ref{Fig. 2}(b)). Furthermore, $\Delta l_c^h$ in Eq. \ref{eq. 1} increases with increasing $l$ due to the longer length of the droplet contact line on top of the electrode elements with longer $l$ (Fig. \ref{Fig. 2}(d)). Accordingly, the electrical trapping force increases with increasing $l$; consequently, $\langle R_{sh} \rangle$ increases with increasing $l$ for a higher value of $U_{rms}$ (Fig. \ref{Fig. 2}(b)). Eq. \ref{eq. 1} provides a reasonable estimate for $\langle R_{sh} \rangle$ under ac-EW with zigzag interdigitated electrodes; e.g., the experimental values of $\langle R_{sh} \rangle$ for $l=$3000 $\mu$m, at $U_{rms}=100$ V  and 150 V, are 1178 $\mu$m and 1333 $\mu$m respectively, while the corresponding values estimated from Eq. \ref{eq. 1} are $\sim1038$ $\mu$m and $\sim1351$ $\mu$m respectively (considering $\Delta \cos \theta_0 \approx 0.14$, $\beta \sim 8.5 \times 10^{-6}$ F$/$(N$\cdot$m), $\alpha \sim 0.8$, $\Delta l_c^h\sim O(2\langle R_{sh} \rangle)$). Finally, the increasing value of $\langle R_{sh} \rangle$ due to the electrical trapping effect results in lower shedding rate for the zigzag interdigitated electrodes, as compared to the straight interdigitated electrodes (Fig. \ref{Fig. 2}(c)).                 

\begin{figure} [ht]
	\includegraphics*[width=\columnwidth]{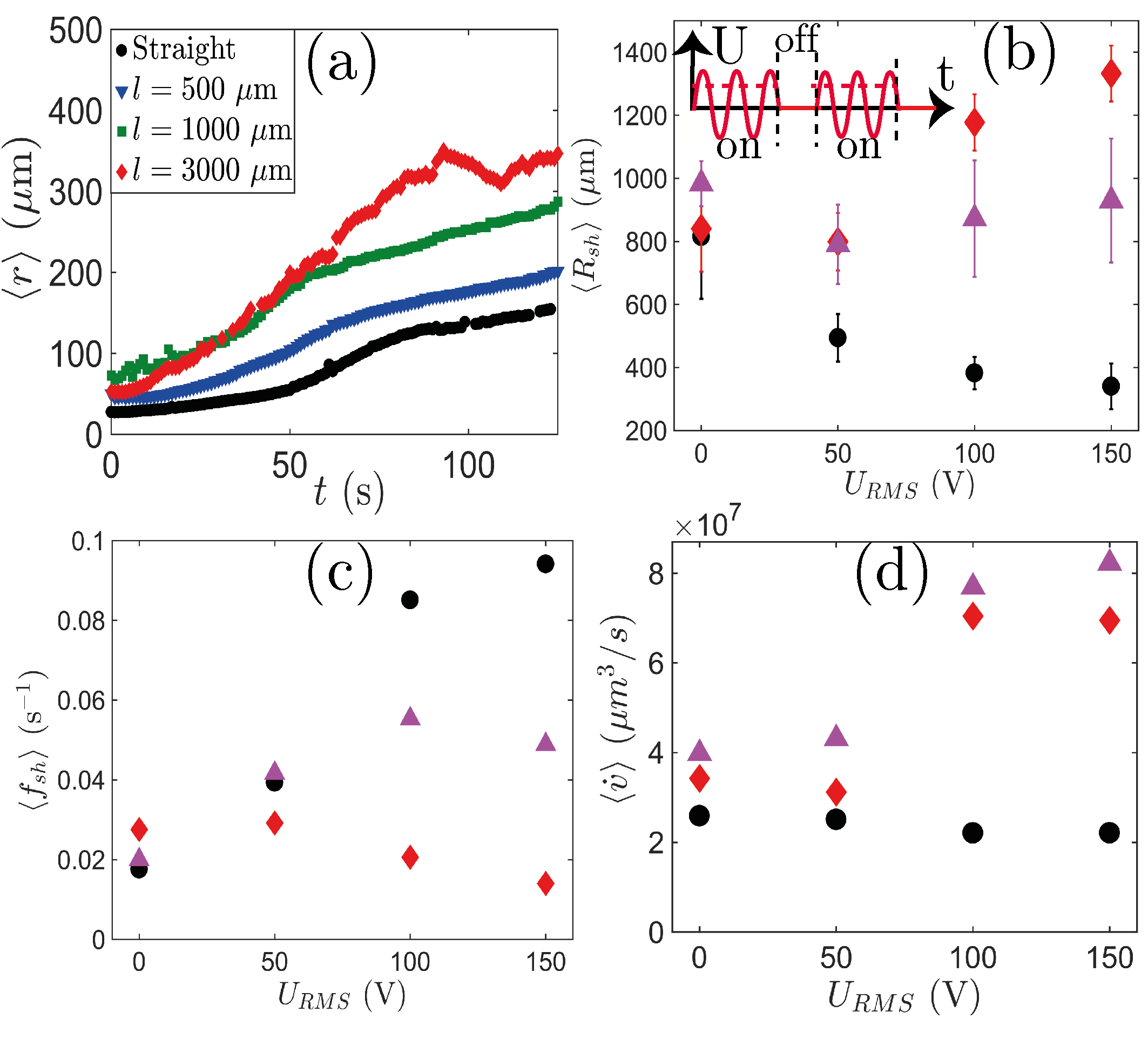}
	\caption{\label{Fig. 3} (a) Temporal variations of the area-weighted mean radius $\langle r \rangle$ of the condensate drops under ac-EW ($U_{rms}=150$ V) with different electrode designs. (b) Variations of the average shedding radius $\langle R_{sh} \rangle$ with the applied voltage $U_{rms}$ under continuous ac-EW (circles: straight interdigitated electrodes; diamonds: zigzag interdigitated electrodes with $l=3000$ $\mu$m), and under intermittent ac-EW (triangles: identical zigzag interdigitated electrodes). The intermittent ac-EW is achieved by switching the applied sinusoidal voltage on (50 s) and off (10 s) as shown in the inset. (c) Variations of the average shedding rate $\langle f_{sh} \rangle$ with $U_{rms}$ for continuous and intermittent ac-EW. (d) Variations of the average volumetric condensate removal rate $\langle \dot{v} \rangle$ with $U_{rms}$ for the different ac-EW conditions. The symbols in (c) and (d) represent the identical EW conditions defined in (b).}
\end{figure}

Ac-EW with zigzag interdigitated electrodes indeed leads to faster growth of condensate drops at earlier stages due to the electrically induced droplet sweeping, as shown by the temporal variations of the area-weighted average radius of the condensate drops $(\langle r \rangle(t)= \Sigma r^3/\Sigma r^2)$ (Fig. \ref{Fig. 3}(a)). $\langle r \rangle$  increases with increasing $l$  (Fig. \ref{Fig. 3}(a)) because the drops sweep larger areas of the surface and thereby undergo more coalescence before getting trapped, as can be seen in Fig. \ref{Fig. 1}(b). However, this faster growth does not translate into a higher shedding rate because of the electrostatic trapping effect, as described before. The effect of electrostatic trapping can be attenuated by applying $U_{rms}$ intermittently instead of continuously, as shown in the inset in Fig. \ref{Fig. 3}(b). During the voltage-on phases, the sweeping and enhanced coalescence of drops promote faster growth towards a radius $\sim \lambda_c$ (note that in absence of EW, $\langle R_{sh} \rangle \sim O(\lambda_c)$). Subsequently, the voltage-off phases facilitate the gravity-driven shedding of the sufficiently big condensate drops by turning off the electrical traps. Fig. \ref{Fig. 3}(b) (red diamonds vs. purple triangles) shows that $\langle R_{sh} \rangle$ under ac-EW with zigzag interdigitated electrodes decreases under intermittent ac-EW as compared to continuous ac-EW due to the periodic switching off of the electrical traps; note that the resulting value of $\langle R_{sh} \rangle$ under intermittent excitation is comparable to that obtained without EW ($U_{rms}=0$ V). However, the resulting $\langle f_{sh} \rangle$ increases under intermittent ac-EW as compared to that obtained without EW (Fig. \ref{Fig. 3}(c)) due to the associated faster growth rate of drops (Fig. \ref{Fig. 3}(a)). From an applied perspective, the most interesting performance indicator of a condensation process is the total condensate volume obtained per unit time ($\dot{v}$). We can estimate an average value of $\dot{v}$ corresponding to the different EW-controlled shedding characteristics by simply multiplying $\langle f_{sh} \rangle$ with the average volume of the shedding drops ($\langle \dot{v} \rangle \approx \frac{2}{3} \pi \langle R_{sh} \rangle^3 \times \langle f_{sh} \rangle$) (Fig. \ref{Fig. 3}(d)). In this regard, $\langle \dot{v} \rangle$ remains similar with increasing $U_{rms}$ for ac-EW with straight interdigitated electrodes (Fig. \ref{Fig. 3}(d)); this is because the associated reduction in $\langle R_{sh} \rangle^3$ (Fig. \ref{Fig. 2}(b)) nullifies the increase in $\langle f_{sh} \rangle$ (Fig. \ref{Fig. 2}(c)). However, $\langle \dot{v} \rangle$ significantly increases (almost by a factor of 2) for ac-EW with zigzag interdigitated electrodes; specifically, the case of intermittent ac-EW is relatively advantageous than continuous ac-EW (Fig. \ref{Fig. 3}(d)). This is because in the case of continuous ac-EW with zigzag interdigitated electrodes the significant increase in $\langle R_{sh} \rangle^3$ (Fig. \ref{Fig. 2}(b)) dominates over the reduction in $\langle f_{sh} \rangle$ (Fig. \ref{Fig. 2}(c)) resulting in the significant enhancement in $\langle \dot{v} \rangle$; for the corresponding intermittent ac-EW the increase in $\langle f_{sh} \rangle$ compensates for the reduction in $\langle R_{sh} \rangle$ (Fig. \ref{Fig. 3}(b), (c)). In fact a more accurate estimate of $\langle \dot{v} \rangle$ should also include the condensate drop volumes swept away by the shedding drop, which should lead to even higher net condensation rates. Yet, such an analysis is beyond our current scope.\\

In summary, we have shown that the gravity-driven shedding of condensate drops can be enhanced using ac-EW with structured electrodes. The effect results from a combination of the ac-EW-induced reduction of CAH and the faster growth of the drops induced by enhanced drop motion and subsequent coalescence. It must be also noted here that the methodologies for implementing ac-EW towards achieving enhanced condensate shedding are not restricted to those demonstrated here; different electrode designs and electrical voltage waveforms can be still conceived for further optimization of the shedding characteristics. Finally, the observed ac-EW induced shedding characteristics will definitely alter the associated heat transfer characteristics from those classically observed for dropwise condensation. However, rigorous experiments involving simultaneous measurements of heat transfer and shedding characteristics under different ac-EW conditions are necessary to identify the optimum conditions for enhanced heat transfer. Such an endeavour can definitely be the scope of future research work. \\

See the supplementary material for the movies (S1-S5) of breath figure evolution under different ac-EW conditions, schematic of the experimental setup (Fig. S1), image analysis procedure, discussion on CAH under ac-EW in air (Fig. S2), and discussion on the electrical trapping effect (Fig. S3).\\      
\begin{acknowledgments}
We thank Daniel Wijnperle for his immense help with the preparation of the condensation substrates. We acknowledge financial support by the Dutch Technology Foundation STW, which is part of the Netherlands Organization for Scientific Research (NWO), and the VICI program (grant 11380).
\end{acknowledgments}

\bibliographystyle{apsrev4-1}
\bibliography{ew_shedding_AUG_2018}
\end{document}